\documentclass[aps,prl,twocolumn,showpacs,superscriptaddress, twocolumn, amssymb,showpacs,floatfix,nofootinbib,tighten]{revtex4-2}
\usepackage{amssymb,amsmath,graphicx,color,microtype}
\usepackage{graphicx}
\usepackage{txfonts}

\usepackage{epsf}
\usepackage{epstopdf}
\usepackage{amssymb}
\newcommand{\nc}{\newcommand}
\newcommand\be{\begin{equation}}
\newcommand\ee{\end{equation}}

\usepackage{float}

\nc{\e}{{\bf{e}}}
\nc{\kk}{{\bf{k}}}
\nc{\pp}{{\bf{p}}}

\nc{\bfk}{{\bf{k}}}
\nc{\bfx}{{\bf{x}}}
\nc{\bfp}{{\bf{p}}}

\nc{\eH}{{\epsilon_H}}
\nc{\calP}{{\cal P}}
\nc{\im}{{ \mathrm{Im} } }

\def\apj{Astrophysical Journal}% Astrophysical Journal,Letters}
\def\mnras{Monthly Notes of Royal Astronomical Society}

\def\araa{Annual Review of Astronomy \& Astrophysics}

\def\aap{Astronomy and Astrophysics}     
\def\prd{Phyical Review D}

\begin{document}

\author{Tom~Broadhurst}
\affiliation{Department of Theoretical Physics, University of the Basque Country UPV-EHU, 48040 Bilbao, Spain}
\affiliation{Donostia International Physics Center (DIPC), 20018 Donostia, The Basque Country}
\affiliation{IKERBASQUE, Basque Foundation for Science, Alameda Urquijo, 36-5 48008 Bilbao, Spain}

\author{Jose~M. Diego}
\affiliation{Instituto de F\'isica de Cantabria, CSIC-Universidad de Cantabria, E-39005 Santander, Spain}

\author{George~F.~Smoot}
\affiliation{ IAS TT \& WF Chao Foundation Professor, IAS, Hong Kong University of Science and Technology,
Clear Water Bay, Kowloon, 999077 Hong Kong, China}
\affiliation{Paris Centre for Cosmological Physics, Universit\'{e} de Paris, CNRS,  Astroparticule et Cosmologie, F-75013 Paris, France A, 10 rue Alice Domon et Leonie Duquet,
75205 Paris CEDEX 13, France}
\affiliation{Physics Department and Lawrence Berkeley National Laboratory, University of California, Berkeley,
94720 CA, USA }

\title{A Distant Origin For Magnified LIGO/Virgo Black Holes Implied By Binary Component Masses }

%\author{Tom Broadhurst$^{1,2,3}$, Jose M. Diego$^{4}$, George F. Smoot$^{5, 6,7,8}$}

% affiliations 
% -------------------------------------------------------
%\begin{affiliations}

%\item {Department of Theoretical Physics, University of The Basque Country UPV/EHU, E-48080 Bilbao, Spain}
%\item{Donostia International Physics Center (DIPC), 20018 Donostia, The Basque Country}
%\item {IKERBASQUE, Basque Foundation for Science, E-48013 Bilbao, Spain}
%\item {Instituto de F\'isica de Cantabria, CSIC-Universidad de Cantabria, E-39005 Santander, Spain}
%\item {IAS TT \& WF Chao Foundation Professor, IAS, Hong Kong University of Science and Technology,
%Clear Water Bay, Kowloon, 999077 Hong Kong} 
%\item {Paris Centre for Cosmological Physics, APC, AstroParticule et Cosmologie, Universit\'{e} Paris Diderot,
%CNRS/IN2P3, CEA/lrfu, %Observatoire de Paris,  
%Universit\'{e} Sorbonne Paris Cit\'{e}, 10, rue Alice Domon et Leonie Duquet,
%75205 Paris CEDEX 13, France}  
%\item {Physics Department and Lawrence Berkeley National Laboratory, University of California, Berkeley, 94720 CA, USA}
%\item {Physics Department and Energetic Cosmos Laboratory, Nazarbayev University, Astana, Kazakhstan}
%\end{affiliations}

%\begin{linenumbers}
% abstract
% -------------------------------------------------------

%\boldmath

\begin{abstract}

The primary and secondary masses of the binary black holes (BBH) reported by LIGO/Virgo are correlated with a narrow dispersion that appears to increase in proportion to mass. The mean binary mass ratio  $1.45\pm0.07$ we show is consistent with pairs drawn randomly from the mass distribution of black holes in our Galaxy. However, BBH masses are concentrated around $\simeq 30M_\odot$, whereas black holes in our Galaxy peak at $\simeq 10M_\odot$. This mass difference can be reconciled by gravitational lensing magnification which allows distant events to be detected with typically $z\simeq 2$, so the waveform is reduced in frequency by $1+z$, and hence the measured chirp masses appear 3 times larger than their intrinsic values. This redshift enhancement also accounts for the dispersion of primary and secondary masses, both of which should increase as $1+z$, thereby appearing to scale with mass, in agreement with the data. Thus the BBH component masses provide independent support for lensing, implying most high chirp mass events have intrinsic masses like the stellar mass black holes in our Galaxy, coalescing at $z>1$, with only two low mass BBH detections, of $\simeq 10M_\odot$ as expected for unlensed events in the local Universe, $z\simeq 0.1$. This lensing solution requires a rapidly declining BBH event rate below $z<1$, which together with the observed absence of BBH spin suggests most events originate within young globular clusters at $z>1$, via efficient binary capture of stellar mass black holes with randomly oriented spins.

\end{abstract}
%\end{boldmath}
% main text

\maketitle

% -------------------------------------------------------

%\section{Introduction}

  Normal stellar black holes in our Galaxy, found in binary configurations \cite{Remillard}\cite{El-Badry}\cite{Stanway}\cite{Simon-Diaz}, have a relatively narrow log-normal mass distribution peaked at $\simeq 8M_\odot$ that can be detected by LIGO/Virgo only in the nearby universe, within $z\simeq 0.25$, see Figure~1. It is striking that the two lowest BBH chirp mass events do fall squarely within this expected narrow range of mass and distance, as shown in Figure 1. In contrast the other 16 black holes, corresponding to the 8 reported BBH events of higher chirp mass, are clearly peaked at $\simeq 30M_\odot$, well beyond the observed range of stellar black holes, as shown in Figures 1 \&\ 2.  This high mass peaked distribution of BBH events is incompatible with an extended power-law mass distribution of binaries initially defined from the first few BBH events\cite{Abbot_2016} and was employed in early gravitational lensing calculations\cite{Ng} for which lensed events are swamped in number by a shallow power-law tail of unlensed events.

%\FloatBarrier
\begin{figure*}[ht]
 \center
 \vspace{-1pt}

  \includegraphics[width=\textwidth]{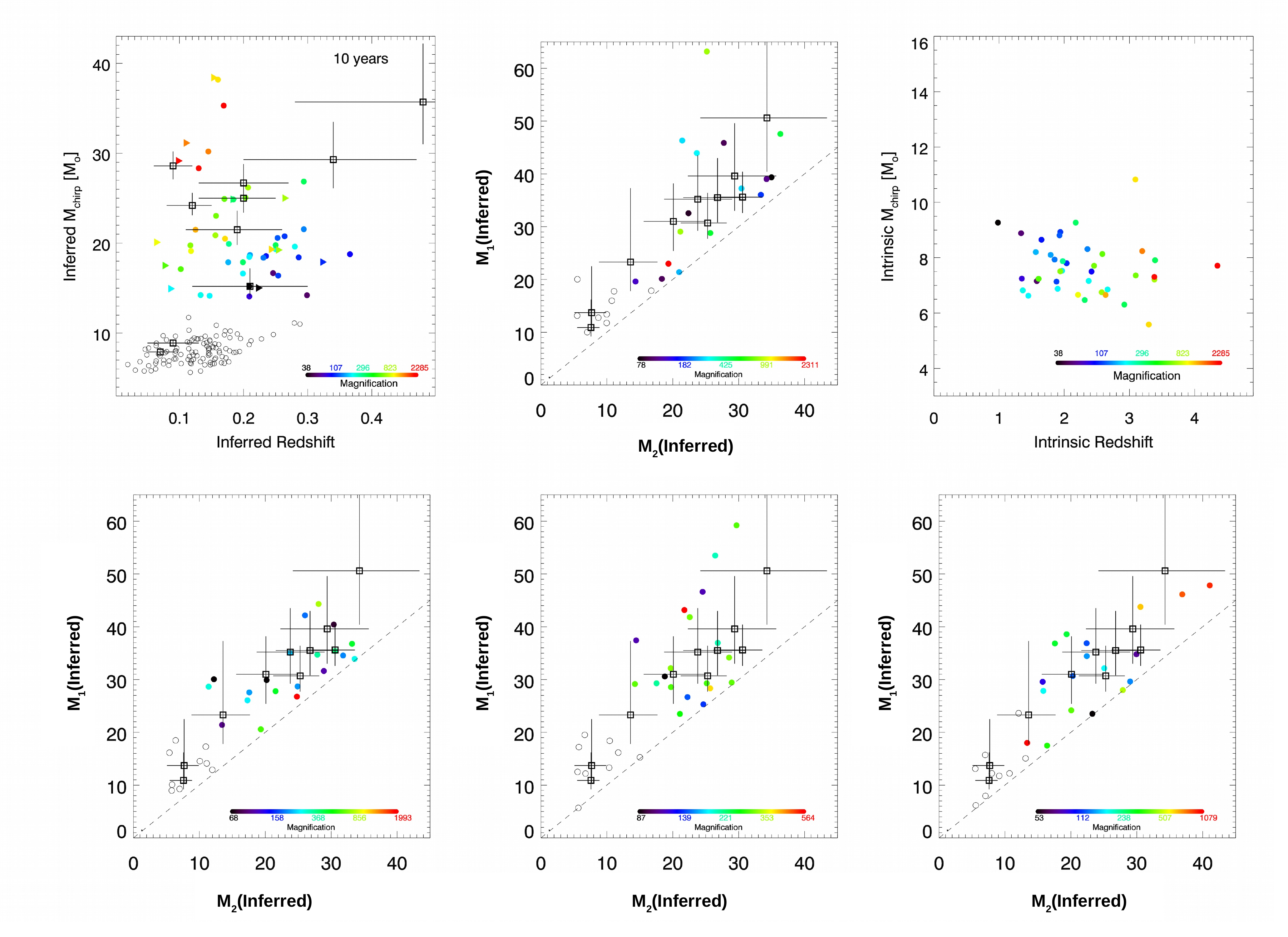}
  \vspace{-4pt}
\caption{{\bf Comparison of reported events with our gravitational lensing predictions.}  The open and coloured circles are our unlensed and lensed event predictions, respectively, where
the colour indicates magnification. Colored triangles indicate a repeat event where two events are detected. The upper left panel includes all the reported chirp masses and distances compared with our lensing model, where the chirp mass is corrected for the relatively small redshift inferred without correction for the strain magnification, like the data. It can been seen that the majority of observed events are consistent with being magnified, with two observed events lying within the expected unlensed region, for which the reported distances can be relied on. The upper middle panel shows how this same lens model clearly accounts for the observed  "V-shaped" distribution of primary and secondary masses visible in the data, demonstrating that the majority of events are redshifted to apparent masses that are 2-3 times larger than their intrinsic values drawn from the relatively narrow mass function of black holes known in our Galaxy. In contrast, a broad mass function would be expected to populate the region of high $M_1$ and low $M_2$in this plane. The upper right panel shows the intrinsic redshifts and chirp masses of the predicted events in the left panel. The lower row shows three more realizations of the mass plane to illustrate the relatively small variance we predict for unlensed events compared to the wider dispersion of higher redshift lensed events.}
\label{r0-value1}
\end{figure*}

 We have shown that the surprisingly high chirp masses detected to date can be entirely explained by gravitational lensing when using the known mass distribution of black holes in the Milky Way and local group found in binary configurations, which does not extend above $\simeq 20M_\odot$. Such stellar mass gravitationally lensed BBH events are predicted to be currently detectable in the redshift range $1<z<3$, with chirp masses dominated by the cosmological redshift of the waveform by $1+z$, that exactly mimics a higher chirp mass binary \cite{Broadhurst2018}. This is because gravitation is scale free, so binaries with masses $(m_1, m_2)$ in the rest frame are indistinguishable by frequency dependence from masses $(\frac{m_1}{1 + z} , \frac{m_2}{ 1 + z})$ at redshift $z$, as the cosmological time dilation ensures the waveform is shifted by precisely the necessary amount to preserve the chirp shape and frequency, by the same factor of $1+z$ for both the primary and secondary masses. Hence, there is the possibility of a much larger downward correction, by $1+z$, than has been hitherto applied. Furthermore, the lensing magnification of the waveform amplitude is not directly measured and so there is considerable degeneracy possible between gravitational magnification, ($\mu$), and luminosity distance $D_L$ \cite{Wang} that may cause the luminosity distance to be revised upward by a factor of $\sqrt{\mu}$, allowing an indirect estimate of the source redshift, $z_{ind}$

\begin{equation}
D_L(z_{ind}) =\frac{{\rm M}_{chirp, z}^{5/6}}{h(t)} ~ F(t,{\rm M}_{obs},\Theta) = \frac{D_L(z)}{\sqrt{\mu}}
\label{Eq_ht}
\end{equation}
where ${\rm M}_{chirp, z} = (1+z) (m_1 m_2)^{3/5} /(m_1 + m_2)^{1/5}$ is the redshifted (observed) chirp mass and $F(t,{\rm M}_{obs},\Theta)$ combines the angular sky sensitivity, orbital inclination, spin and polarization of the binary source\cite{Finn} and its distribution is numerically estimated with a $\simeq 40\%$ dispersion\cite{Ng,Abbot_2016}. 

Using equation~1 we find that lensing of normal stellar mass black hole binaries can enable the detection of large, apparent BBH chirp masses shown in Figure~1 (coloured points) because these lie well above the chirp mass range expected for an unlensed population of stellar mass BBH events as shown in Figure~1 (open circles), that are detectable to only $z\leq 0.25$ with the current sensitivity of LIGO/Virgo. This distinction is analogous to the brightest infrared galaxies detected in large sky surveys that are known to be predominantly lensed by intervening galaxies, forming obvious Einstein rings, with a mean radius of $\simeq 0.85"$  \cite{Wardlow,Negrello,Bussmann}
indicating that lensing has boosted the flux of distant background infrared galaxies (typically at $z\simeq 2$) so much they exceed the fluxes of the brightest unlensed galaxies at lower redshift. This effect is stronger for small sources like a BBH events as they may be projected ``closer" to a lensing caustic where the magnification diverges.

 Our predictions are made simply with an input mass distribution of binaries drawn randomly from the known log-normal stellar black hole mass function. We then lens these binaries with the Universal form of the high magnification tail of fold caustics, $\propto \mu^{-3}$, normalised to the known optical depth of galaxy lensing. Finally we adopt an exponentially declining BBH event rate since redshift of $z=2$ as this simple assumption matches well the observations, predicting an approximate $3/1$ ratio of lensed to unlensed events, as shown in Figures 1\&2, corresponding to an e-folding time of $\simeq 1 Gyr$. This  evolution timescale is the most free parameter given the very uncertain origin of BBH events both at low and high redshift. Details of the model can be found in \cite{Broadhurst2018} and \cite{Diego2020}, referred to as the BDS model.

 We can also examine the 20 individual primary and secondary black hole masses reported by LIGO/Virgo\cite{Abbott1+2}, for comparison with the above lensing model, shown in Figure~1. It can be seen how well our simple lensing model reproduces the published component masses data, both in terms of the binary mass ratio, which is observed to be independent of mass and also in terms of the ``V-shaped" dispersion in their apparent masses, $\sigma_m$. We predict a relatively small dispersion of primary and secondary masses at the low mass end, corresponding to unlensed sources at a mean redshift of $z\simeq 0.1$. This dispersion should become approximately 3 times larger at higher masses, because we predict the origin of this increasing dispersion is simply the $1+z$ expansion of their apparent masses, with a mean source redshift, so our predicted dispersion for sources at $z_s$, with dispersion $\sigma_m$ simply scales as: 
 \begin{equation}
    \sigma_{m}=\sigma_{un}{(1+z_{s})/(1+\overline{{z}}_{un})}
    \end{equation}    
where $\sigma_{un}$ is the dispersion of the unlensed events of mean observed redshift, $\overline{z}_{un}$,
which we predict to be $\overline{{z}}_{un}=0.1$, that is only 10\% larger than the rest frame intrinsic dispersion. This simple $1+z$ scaling is consistent with the distinctive mass plane distribution shown in Figure~1b, for which no new free parameters are required, other than those we have adopted for the chirp mass-distance plane in Figure~1a and the mean measurement error is incorporated as a Gaussian of width $\simeq 5M_\odot$. The mean predicted binary mass ratio $M_1/M_2$ is $1.45\pm0.05$, in very good agreement with the data. This encouraging agreement simply says that the apparent correlation between primary and secondary masses is not intrinsic to the distribution of  binary black hole masses but is the result of the cosmological expansion of their waveforms expected in the context of lensing, over the range $1<z_s<4$. 

Soon, Run 3 data release,  three times as many events will make it possible to define accurately the redshift distribution, with little model dependence by using the width of the dispersion in the component mass plane and reading number off the number density of BBH events along the mass correlation. The lensed and unlensed populations in the component mass plane should be readily distinguishable we predict with our lensing model as shown in Figure~2, dividing relatively cleanly into 2 peaks lying above and below a chirp mass of $\simeq 15M_\odot$. 

%\FloatBarrier
\begin{figure}[ht]
 %\center
 \vspace{-1pt}
  \includegraphics[width=0.5\textwidth]{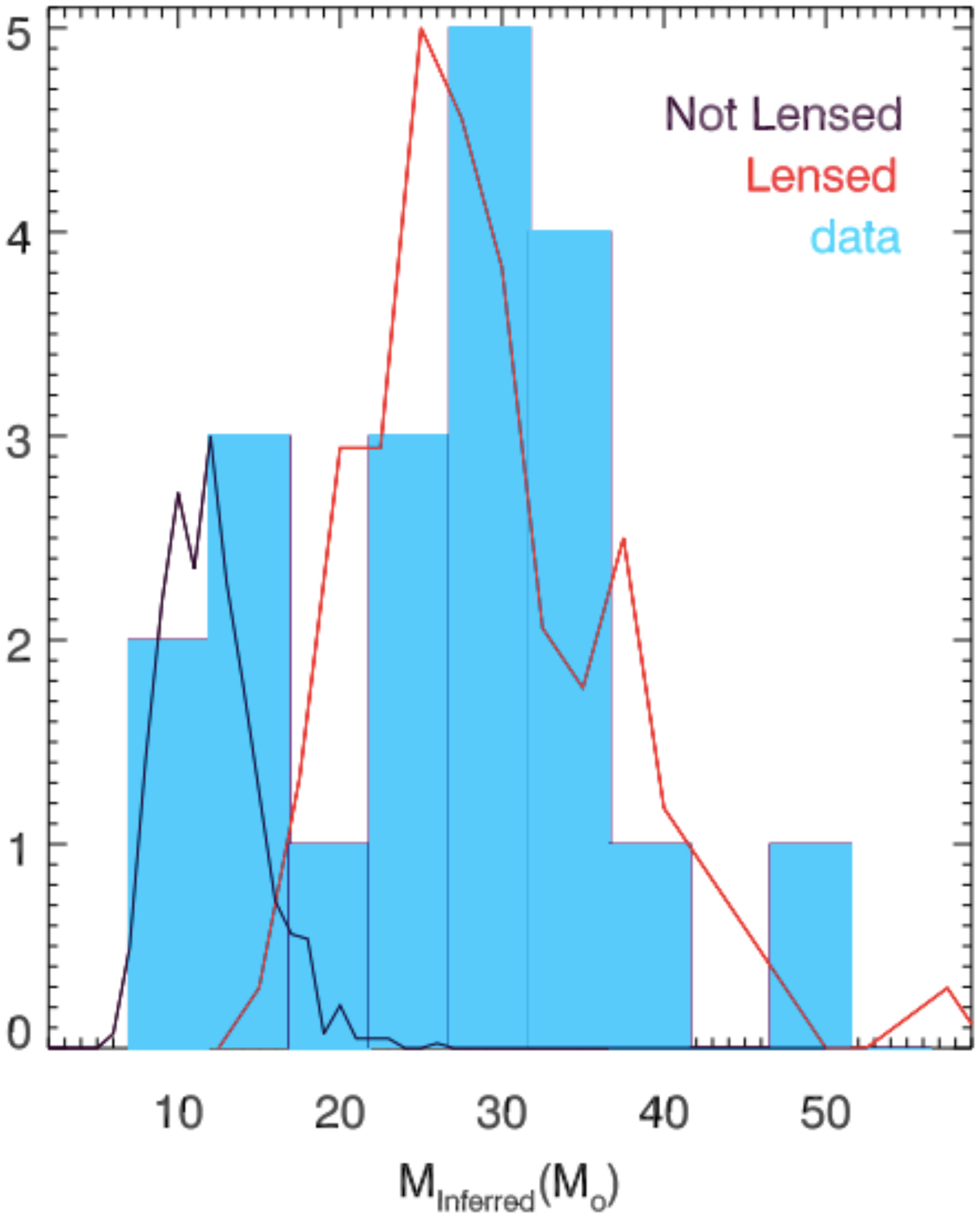}
  \vspace{-1pt}
\caption{{\bf Distribution of binary black hole masses} The histogram includes all 20 primary and secondary BH masses of the 10 reported BBH events. The red curve shows our lensing prediction and the black curve is our prediction for unlensed events. Both predicted curves have characteristic peaks because they sample the relatively narrow log-normal distribution of the known stellar black holes in our Galaxy. The predicted peak redshifts are $z\simeq 0.1$ and $z\simeq 2$ for the unlensed and lensed events respectively, forming a bimodal distribution of predicted BBH masses, similar to the apparent bimodality of the data.}
\label{r0-value2}
\end{figure}
Amongst the magnified population of BBH events we have anticipated the existence of repeated, multiply-lensed events for a significant minority $\simeq 10\%-20\% $ of the BBH mergers. Given the Earth-rotating angular sensitivity of LIGO/Virgo\cite{Broadhurst2018} means that at any one time detection is effectively limited to a overhead (and underfoot) band of the sky that would typically miss the counter image of any given BBH detection, as highly magnified pairs of lensed images generated by galaxy lensing are known to be typically separated by hours to days. We estimate this with our simulation in Figure~1, based on lensing by the intervening population of galaxies,  49 detections  over 5 years of observations, only 9 pairs of events are expected to be close enough in time and magnitude to be potentially detected (coloured triangles in Figure~1 top left), but only about 10\%\ will be due to the rotating angular dependence of the detection limit and a requirement of detection by coincidence in at least two detectors. 

Non-detection is also a problem even for rare pairs of events more closely separated in time, of less than an hour, for which there would be no significant sensitivity variation, because it has been empirically demonstrated that the relative flux of close image pairs, typically differ by a factor 2 in brightness and sometimes more\cite{SNlens}, implying substructure is universal on sub-arcsecond scales\cite{Treu2019} as normal smooth galaxy lensing potentials cannot generate such large magnification differences, and neither can stellar microlensing be invoked for sufficiently extended sources\cite{Treu2019}. A promising explanation for these long standing "flux anomalies" is the inherently granular density field predicted for light bosonic "Wave-Dark-Matter"\cite{Schive2014} due to self-interference. We emphasise that estimates of the incidence of strongly lensed GW events based on smooth lens models\cite{McNowt} are empirically inappropriate.

We look forward to examining the waveforms of over 30 new BBH events when released, but in the meantime is has been asserted, in the absence of published waveforms, that two of these BBH events are not lensed despite being close on the sky and separated in time by only 20 minutes, as they are apparently not sufficiently coincident on the sky \cite{Singer}. We nevertheless would interpret one or both of these as unrelated lensed sources, should it transpires that their chirp masses lie in the lensing region predicted by our model, exceeding $\simeq 15M_\odot$. We have already highlighted such a viable case in the published data separated by 5 days and coincident in position and phase \cite{Broadhurst2019}, with more definitive examples expected,  particularly those that can be triangulated in position by combining LIGO and Virgo, with identical waveforms, with consistent phase, and polarisation.

% Figure for the configuration and layout.
%\includegraphics[]{detmap.png}

%includegraphics[]{Optimal_Networks.png}

Our lensing interpretation finds a rate of BBH coalescence that is highest at $z\simeq 1.5$ with an intrinsic rate of $\simeq 10^{4}yr^{-1}Gpc^{-3}$, and a steep decline thereafter to about $\simeq 10yr^{-1}Gpc^{-3}$ by $z\simeq 0.1$ \cite{Broadhurst2018}, corresponding to an e-folding time of $\sim 1$ Gyr, to match the relatively small proportion of 2 to 3 unlensed BBH events out of a total of 10 reported BBH events, as shown in Figure~1. Although the peak we infer at $z\simeq 1.5$ is approximately coincident with the maximum rate of cosmological star formation\cite{Madau}, the evolution of the star formation rate is shallower than we infer for BBH events, dropping by only one order of magnitude since its peak $z\simeq 2$ to the present day, compared with a decline of $\simeq 3$ orders of magnitude for BBH events in the context of lensing.

Such strong evolution may point to BBH events originating within globular clusters that are known to form predominantly at $z>1$ and are long understood to provide an efficient environment for the capture of stellar mass binaries\cite{Sig,Banerjee}. In particular, metal rich globular clusters that are common in massive early type galaxies have estimated ages corresponding to $1<z<4$\cite{Forbes2015}, in agreement with the accurately dated metal rich globular clusters in our Galaxy and direct detections of stellar mass black holes in binaries have been uncovered in nearby globular clusters \cite{NGC3201}. Furthermore, we may also favour a capture origin from our conclusion that the BBH distribution can be reproduced well by pairs drawn randomly from the known stellar mass black hole distribution, and for which spins should be unaligned, allowing an understanding of the puzzling lack of BBH spin reported for the BBH events \cite{Farr}.

The lack of spin has motivated coalescence calculations within dense star clusters for which high mass BBH events may be generated after repeated BH merging. However, a wide spread in BBH masses is predicted, dominated by lower mass stellar black holes $\simeq 10M_\odot$ with a relatively rare massive tail\cite{Ugo}. Whereas lensing in this dense cluster context reduces the need for repeated BH merging, as the large cosmological redshift enhances the chirp masses well beyond their intrinsic stellar values. We note that an origin for massive BBH events in terms of primordial black holes comprising the dominant dark matter\cite{Bird,Carr} seems disfavoured given the modest level of microlensing of several newly discovered distant stars lensed by large columns of dark matter on the Einstein radius of massive lensing clusters \cite{Kelly,Venumadhav,Diego,Oguri}. A smaller population is allowed by these lensed stars but leaves over 90\% of the dark matter to be explained in other ways.

The population of metal poor globular clusters is widely estimated to be 2 Gyrs older than the metal rich population, corresponding to significantly higher formation redshifts, with $z > 5$, that may conceivably provide a population of even higher chirp mass BBH events, exceeding $\simeq 60 M_\odot$, but these should be relatively rare, requiring extreme magnification to counter the large luminosity distance at such high redshifts, given the current LIGO/Virgo sensitivities.

We can look forward to the threefold increase in BBH events already detected by the LIGO/Virgo team, allowing a much more accurately defined chirp mass-distance plane and binary component masses. We predict lensed events will continue to dominate GW detections for the foreseeable future, until increased sensitivity can provide sufficient volume for the unlensed population of $\simeq 10 M_\odot$ stellar mass black holes to dominate detections at lower signal strength and lower redshift. This division between relatively local and much more distant events is a bonus for gravitational wave astronomy, allowing the evolution of stellar black hole properties to be compared over a Hubble time.

%\end{linenumbers}

%\clearpage

\end{document}